\documentclass[10pt,journal]{IEEEtran}

\usepackage[pdftex]{graphicx}
\usepackage{pgf}
\usepackage{amsmath}

\usepackage{algorithm}
\usepackage{algorithmic}

\usepackage{amssymb}
\usepackage{xcolor}
\usepackage{subcaption} 
\usepackage{booktabs} \usepackage{tabularx}
\usepackage[bookmarks=false,hidelinks]{hyperref}

\everymath=\expandafter{\the\everymath\textstyle}

\newcommand{\algorithmname}{\texttt{Semantica}}

\begin{document}

\title{Semantica: Decentralized Search using a LLM-Guided Semantic Tree Overlay} 

\author{%
    \IEEEauthorblockN{Petru Neague\IEEEauthorrefmark{1}, Quinten Stokkink\IEEEauthorrefmark{1}, Naman Goel\IEEEauthorrefmark{2}, Johan Pouwelse\IEEEauthorrefmark{1}}

    \IEEEauthorblockA{\IEEEauthorrefmark{1}Delft University of Technology
    \\\{p.m.neague, q.a.stokkink, j.a.pouwelse\}@tudelft.nl}

    \IEEEauthorblockA{\IEEEauthorrefmark{2}University of Oxford
    \\naman.goel@cs.ox.ac.uk}
}

\maketitle

\begin{abstract}
Centralized search engines are key for the Internet, but lead to undesirable concentration of power.
Decentralized alternatives fail to offer equal document retrieval accuracy and speed.
Nevertheless, Semantic Overlay Networks can come close to the performance of centralized solutions when the semantics of documents are properly captured.
This work uses embeddings from Large Language Models to capture semantics and fulfill the promise of Semantic Overlay Networks.
Our proposed algorithm, called Semantica, constructs a prefix tree (trie) utilizing document embeddings calculated by a language model.
Users connect to each other based on the embeddings of their documents,  ensuring that semantically similar users are directly linked.
Thereby, this construction makes it more likely for user searches to be answered by the users that they are directly connected to, or by the users they are close to in the network connection graph.
The implementation of our algorithm also accommodates the semantic diversity of individual users by spawning ``clone'' user identifiers in the tree.
Our experiments use emulation with a real-world workload to show Semantica's ability to identify and connect to similar users quickly.
Semantica finds up to ten times more semantically similar users than current state-of-the-art approaches. At the same time, Semantica can retrieve more than two times the number of relevant documents given the same network load. We also make our code publicly available to facilitate further research in the area.
\end{abstract}



\section{Introduction}
Search engines like Google are the cornerstone of how individuals share and discover information on the Internet.
Estimates of monthly website traffic in 2024 for \texttt{google.com} range from 80 billion to 135 billion visits~\cite{statista_google, semrush2024google}. 
The centralized nature of widely-used Web technologies has raised a number of concerns related to concentration of power, loss of individual autonomy, privacy, governance, etc~\cite{zuboff2023age, stucke2022breaking}.
As an alternative, decentralized systems such as peer-to-peer systems have been built~\cite{rodrigues2010peer}, but they have seen limited adoption due to a range of technical and non-technical factors~\cite{Baquero2023}.
In this work, we investigate an exciting opportunity presented by the state-of-the-art pre-trained large language models (LLMs)~\cite{min2023recent} to significantly improve the scalability, accuracy, and ease of implementation of systems for \emph{decentralized semantic search}. 

Centralized search engines allow users to quickly discover relevant information from the vast amount of publicly available information.
However, search is technically challenging to implement in the absence of any central entity to index all the information available in a network and serve the queries of users.
This problem has received significant attention in the past, with ideas such as structured overlay networks proposed as potential solutions.
Tried-and-tested solutions exist for models like Distributed Hash Tables (DHTs) and Publish-Subscribe~\cite{vansteen2018distributed} to serve content distribution and key-value-based retrieval. 
However, a decentralized solution towards serving text queries with semantically related results has not yet seen wide adoption. 

\begin{table}
\caption{Characteristics of related retrieval techniques.}
\label{tab:existingwork}
\centering
\setlength{\tabcolsep}{2pt}
\begin{tabular}{p{0.32\linewidth} c c c c c}
\toprule
\textbf{Approach} & \textbf{Semantic} & \textbf{Predictive} & \textbf{Distributed} & \textbf{Training} \\ 
 & \textbf{Search} & \textbf{Cache}  &  & \textbf{Not Required} \\ 
\midrule
 \noindent\parbox[c]{\hsize}{Centralized Search Index (e.g. DSI~\cite{tay2022transformer})} & \checkmark & $\times$  & $\times$ & $\times$ \\[6pt]
De-DSI~\cite{neague2024dsi} & \checkmark & $\times$  & \checkmark & $\times$\\[3pt]
Chord DHT~\cite{stoica2003chord} &  $\times$ & \checkmark & \checkmark & \checkmark \\[2pt]
Graph Diffusion~\cite{giatsoglou2022graph} & \checkmark & $\times$ & \checkmark & \checkmark\\[2pt]
\textbf{Semantica} (This work) & \checkmark & \checkmark & \checkmark  & \checkmark\\
\bottomrule
\end{tabular}
\end{table}

Losses in document retrieval accuracy and speed are common problems stemming from decentralization.
Decentralized solutions are scalable when they distribute workloads over multiple physical machines.
In general, 
accessing the distributed data of these workloads involves communication over networks.
Because of the required network communication, it is very challenging to build scalable features on top of a decentralized network that perform as well as the features offered by centralized competitors.
Further, any potential solutions have to consider fundamental constraints such as information becoming (temporarily) unavailable due to nodes going offline, new nodes joining dynamically and slow network communication relative to local database access, etc. 

Accuracy and speed losses due to decentralization can be mitigated using semantics.
Semantic overlay networks (SONs) are structured overlay networks that connect users based on semantics.
The idea 
is that semantics have \textit{predictive capacity}: users that operate in the same semantic context are interested in the same data and they should be connected.
Users with a similar history of a certain type of document are likely to continue to share more documents they are both interested in.
It makes sense to first query these users when searching for information and it has been shown that semantic networks indeed offer low latency and high accuracy in retrieved documents, comparable to centralized approaches~\cite{tang2003peer}.
This speedup is highly related to the way DHTs connect users based on shared hashes and in \autoref{tab:existingwork} we provide a high-level view of other document retrieval techniques that are discussed in \autoref{sec:relatedwork}.
However, the speedup is then critically dependent on the ability of users in such networks to capture the semantics of data.

Capturing the semantics of data is referred to as Latent Semantic Indexing~\cite{dumais1988using} or just Latent Indexing~\cite{salakhutdinov2009semantic}.
The latent index of a user is typically calculated using the mean of all documents accessed by a user.
Previously, 
two common options were available and utilized, 
but (i) decomposition from words to vectors (e.g., TF-IDF~\cite{tang2003peer}) did not sufficiently capture semantic context and (ii) the alternative of Natural Language Processing had exponential time complexity~\cite{berger1996maximum}.
LLMs present an opportunity to revisit this issue.
A pre-trained LLM captures both semantic context and offers low latency, making it a promising technology to deliver on the promise of networks based on semantics.
Rather than constructing latent indices using human-interpretable \textit{words}, indices can be calculated using LLM \textit{embeddings}.



Work on navigating embeddings to retrieve documents has thus far not considered the possibility of a one-to-one mapping to decentralized networks.
This is a missed opportunity.
For example, in recent work, DSI~\cite{tay2022transformer} introduces ``Semantically Structured Identifiers'' to train prefix trees (tries) for efficient search.
For those familiar with computer networks, the similarity to the tries used in DHTs is striking.
In De-DSI~\cite{neague2024dsi}, a first step toward decentralization is taken using a decentralized-ensemble learning approach, but this approach still requires (online) learning periodically to update the index.
In this work, we move beyond single-purpose training and show that pre-trained general-purpose LLM models are sufficient to create decentralized semantic search using trees (i.e., tries).

We present the first predictive decentralized document search algorithm, \algorithmname{}, to fully exploit recent advances in LLM technology.
Our idea is to create a \textbf{structured semantic overlay network using LLMs} for Latent Semantic Indexing.
This tree is maintained and updated in a peer-to-peer (semantic) overlay network.
Our contributions are as follows.

\begin{itemize}
    \item In \autoref{sec:problem} defines the operational setting, aims and assumptions of our work.
    \item How our \algorithmname{} algorithm constructs and maintains a decentralized semantic tree data structure is presented in \autoref{sec:design}.
    \item The implementation of \algorithmname{}, its emulation using real-world workloads, and its performance metrics are given in \autoref{sec:impperformance} and subsequently discussed in \autoref{sec:resultsdiscussion}.
    \item The algorithmic complexity of \algorithmname{} and data structure properties are derived in \autoref{sec:complexity}.
\end{itemize}

\section{Setting}
\label{sec:problem}

Decentralized semantic search aims to locate relevant documents in a large, distributed network without relying on a central index or coordinator. We consider a system composed of \(N\) autonomous users, each storing a subset of the global document set. The subsets of documents at individuals users are not mutually exclusive. 
A user issues a query (i.e., a textual description) that captures the desired semantic content that they wish to find. The challenge is to route and resolve this query purely through peer-to-peer interactions, in a manner that remains both efficient and accurate at scale.

In this setting, no single user possesses a global index of where documents reside. Instead, each node has only local knowledge: it knows its own documents and maintains a limited ``neighbor list'' of other nodes. For the purposes of this paper we assume that users have \emph{perfect search} on their own device. This means that if they have the single document that the query is looking for, they retrieve it 100\% of the time, and if they don't, then they acknowledge that they don't have it 100\% of the time with no chance of mistakes. We make this assumption as Semantica aims to optimize query-routing and we leave local search out of scope. Network addresses can change over time, and nodes may join or leave at will. Existing decentralized overlays (e.g., \ DHTs~\cite{stoica2003chord}, Publish-Subscribe approaches~\cite{vansteen2018distributed}) typically fall short when dealing with fuzzy or semantic-based queries, where straightforward hashing of content keys is inadequate to find query relevant documents.

\begin{figure}
    \centering
    \begin{subfigure}{0.45\textwidth} 
        \includegraphics[width=\textwidth]{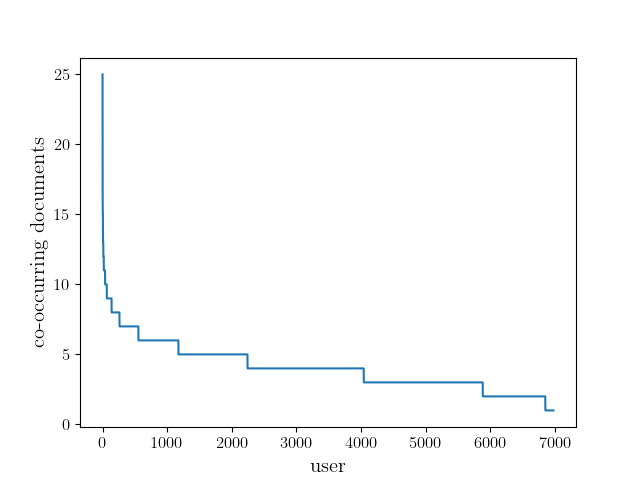}
        \caption{Maximum document co-occurrence with another user per user, sorted by users with the highest co-occurrence first.}
        \label{fig:user_similarity_aol4ps_filtered_dataset}
    \end{subfigure}
    \hfill 
    \begin{subfigure}{0.45\textwidth} 
        \includegraphics[width=\textwidth]{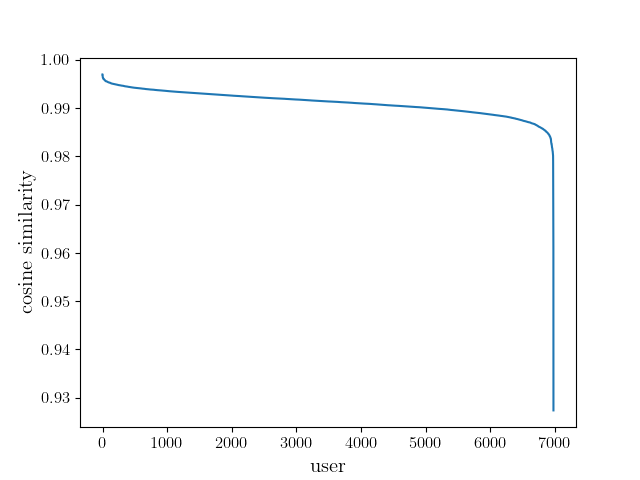}
        \caption{Maximum cosine similarity of mean-document-embedding with another user per user, sorted by users with the highest similarity first.}
        \label{fig:cosine_sim_aol4ps_filtered_dataset}
    \end{subfigure}
    \caption{User similarity in the AOL4PS data set.}
\end{figure}

We assume that users have overlap, i.e., co-occurrence, in the documents that they store.
To show that this is a realistic assumption, we determine the co-occurrence of documents and the cosine similarity of users, sampled from a real-world data set.
We evaluate the AOL4PS~\cite{guo2021aol4ps} data set, consisting of 187\,521 websites accessed by 12\,907 users over the course of three months in 2006, totaling 1\,339\,101 queries.
In \autoref{fig:user_similarity_aol4ps_filtered_dataset}, we show that 6\,978 users have at least one co-occurring document (i.e., website) with another user.
Secondarily, this implies that contacting the remaining 5\,929 users for any search would not be useful and an efficient search solution should not connect to them.
To show that these documents can be used for semantic search, we calculate the average of the embeddings of the documents accessed by users.
We calculate these document embeddings using BERT~\cite{kenton2019bert}.
In \autoref{fig:cosine_sim_aol4ps_filtered_dataset}, we show that the maximum cosine similarity between user embeddings is mostly over 0.93.
This means that, for users that do have document co-occurrence, another user exists with high semantic overlap, which our solution leverages.

We propose to utilize the vector embeddings of documents and queries, obtained from pre-trained large language models (LLMs)~\cite{min2023recent}. Vector embeddings are numerical representations of the data such as words and sentences. Every node can compute or obtain embeddings for its documents, for instance by running a local LLM or querying an external embedding service. A node issues a query in the form of an embedding of a textual query that captures the desired semantic content. Our overarching goal is to design an overlay protocol that enables queries to find semantically relevant documents with minimal network communication and computation overhead. Specifically, we seek to (i) support \emph{semantic search}, where nodes match queries based on embeddings rather than exact keys; (ii) maintain a \emph{predictive} topology, where nodes sharing similar content are connected to one another, thus reducing query path length and lookup times; (iii) operate in a fully \emph{distributed} manner, accommodating a large number of nodes; and (iv) avoid the cost and complexity of large-scale training, by leveraging pre-trained or lightly tuned LLMs at each node.

Decentralized semantic search faces obstacles arising from the partial knowledge at each node. Even if a document is highly relevant, it may reside on a remote node that is unknown to the query initiator. Maintaining a network structure that places semantically similar peers near each other can mitigate this problem, but doing so requires continuous adaptation. Users may discover new documents or alter their interests, causing embedding changes. Nodes may also relocate or leave, meaning the overlay must be robust to churn. Furthermore, limited bandwidth and computational resources—especially if embeddings are computed on commodity hardware or mobile devices—place strict constraints on any adaptive protocol.


Our objective is to show that a \emph{predictive} overlay, built directly from each node’s local embeddings, can preserve scalability by limiting reliance on keeping a global state, while remaining flexible enough to handle evolving interests. Such an overlay dynamically places nodes with similar content ``virtually close'' to each other in the network, so that queries naturally travel to relevant peers without exhaustive searching. The resulting system promises to offer a semantic retrieval akin to that of a centralized engine, yet operates entirely in a peer-to-peer environment. 

\section{Related Work}
\label{sec:relatedwork}

Information retrieval is a vast field. For conciseness, we only discuss the methods that are most closely related to the contribution of our work, namely: decentralized retrieval and retrieval based on machine learning techniques.
Hereby, we also omit related work such as database technology, which is necessary to store documents that require retrieval.
Our discussion of related work is centered around the following defining characteristics, also shown in \autoref{tab:existingwork}:

\begin{itemize}
   \item \textbf{Semantic Search}: whether documents can be retrieved from users by utilizing semantics, instead of a fixed unique key.
   \item \textbf{Predictive Cache}: whether network connections are constructed (in a network overlay) such that documents are likely to be available from existing connections.
   \item \textbf{Distributed}: whether documents are (sparsely) distributed among users that can form connections and retrieve documents from each other.
   \item \textbf{Training Requirement}: whether the solution must be trained on a data set in order to meet its performance targets of latency and accuracy, and must be retrained to update its index.
\end{itemize}


Early peer-to-peer search provided key-value matching from content items to users' search queries that consist of keywords.
For example, Napster allowed users to search their indexing servers for songs, based on song title~\cite{carlsson2001rise}, and
KaZaA would later replace servers with peers~\cite{liang2005kazaa}.
This was similar to the keyword-based search of early web page indexers like Google Search~\cite{brin1998anatomy}.
However, generally, to search for information that is sparsely available from different users, \textbf{distributed} search algorithms are required.
In some use-cases, distributed search algorithms are trivialized, when all information should be spread to all users.
For instance, in blockchain networks like Bitcoin~\cite{nakamoto2008bitcoin} and Ethereum~\cite{wood2014ethereum} information is spread to all users and they can use techniques like flooding or gossip~\cite{vansteen2018distributed}. In our work, we do not make assumptions about search being limited to keyword-based search or to trivial use-cases where all information is spread to all users.

Modern structured networks allow efficient search.
The essence of these networks is to \textbf{predict user searches} and connect users to caches of information.
These localized caches appear in most networking research that structures network topologies around its data, e.g., Content-Centric Networking and Information-Centric Networking~\cite{vasilakos2015information}, and Content-Oriented Networking and Data-Oriented Networking~\cite{cho2008content}.
Typically, a variant of the Publish-Subscribe~\cite{vansteen2018distributed} model is used, in which users subscribe to the information (cache) of a publisher.
In our work, we do not assume content comes from known publishers.

In distributed systems research, content is typically addressed by a pre-known hash.
These hashes allow networks to be structured as a tree, or Distributed Hash Table~\cite{vansteen2018distributed}.
For example, DHTs like Chord~\cite{stoica2003chord} or Kademlia~\cite{maymounkov2002kademlia} use these trees to group users based on the hashes of their user identifiers to the hashes of the data they want to search for.
Our work uses a similar approach for efficient search using users that self-organize in tree topologies.
However, we do not assume users know the hash of the data they are looking for, but instead \textbf{only know the semantics} of that data.
For example, a ``red car'' and not a ``Ferrari 250 GT California Spyder SWB (1959)''.

The idea of using semantics in peer-to-peer overlay networks for search stems from the early 2000's~\cite{tang2003peer}.
In their work on Semantic Overlay Networks (SONs), Tang et al. already point out the scalability limits of approaches that depend on (non-semantic) keywords in a DHT, like that of Li et al.~\cite{li2003feasibility}.
The idea of using ontology trees for efficient semantic search is now more than two decades old~\cite{crespo2004semantic}.
However, these works assumed that a reliable taxonomy is available to create such a tree: this is (still) not the case.

In lieu of a taxonomy, Latent Semantic Indexing (LSI) of data can be used.
These indices fulfill ``adjustable representational richness'', ``explicit representation of both terms and documents'', and ``computational tractability for large datasets''~\cite{deerwester1990indexing}, which we freely interpret as a dynamic, scalable and efficient data structure, such as a tree.
The matrix-based algorithm to fulfill the requirements of LSI by Deerwester et al.~\cite{deerwester1990indexing} would later be misconstrued by many citing works as ``the LSI algorithm'', when it really is only one of the many possible implementations.
That algorithm in particular, used singular value decomposition (SVD), which is one of the many techniques for word embedding~\cite{johnson2024detailed}.
Even so, considering semantic information, using Deerwester et al.'s algorithm, provides superior retrieval accuracy to faster word-frequency-based approaches like TF-IDF~\cite{zhang2011comparative}. 
Our work uses LLMs to construct a Latent Semantic Index using a tree.

More recent approaches to capture the semantics of words, or sequences of words, \textbf{require some form of training}.
For example, GloVe~\cite{pennington2014glove} uses on-line learning to create word vectors (embeddings) based on global word-word co-occurrence counts.
Moving beyond word embeddings, early models for sequence-to-sequence applications, like BERT~\cite{kenton2019bert}, were designed to incorporate positional encoding of words using attention, i.e., transformers~\cite{vaswani2017attention}.
Specifically, BERT was designed to handle NLP tasks and require minimal fine-tuning~\cite{kenton2019bert}.
Later models using transformers, LLMs, became better at more generalized tasks~\cite{min2023recent} (and arguably worse at NLP).
However, LLMs require much less computing resources~\cite{rostam2024achieving}, making them a good fit for consumer hardware.
Most importantly, pre-trained LLM models are widely available.
In our work, we leverage pre-trained models without further training.

A recent influential approach in non-distributed document retrieval based on embeddings is Differentiable Search Index (DSI) by Tay et al.~\cite{tay2022transformer}.
In their work, they train a (sequence-to-sequence) transformer model to retrieve documents.
The model uses its semantic understanding of a user query to retrieve a document identifier, which they refer to as ``Semantic String Docids''.
Hereby, the transformer model is used as a form of latent semantic indexing, though the authors do not mention this.
Furthermore, the paper does not touch on its applicability to a decentralized, or distributed, context.
In their De-DSI approach, Neague et al.~\cite{neague2024dsi} do consider such a context.
In their decentralized approach, they fine-tune a global network by exchanging documents with network peers.
However, efficiency at scale is not yet considered and the time complexity increases as the network grows~\cite{neague2024dsi}.
In our work, we use the semantic understanding of sequence-to-sequence models without any further training and our work does scale with the network size as we use a tree data structure.



The state-of-the-art Graph-Diffusion algorithm by Giatsoglou et al.~\cite{giatsoglou2022graph} uses graph diffusion with Personalized PageRank~\cite{gasteiger2018predict} to spread summarized representations of each node’s content across the network. Each node generates a single ``node embedding'' by aggregating embeddings of its local documents. This node embedding, representing the general content of the node, is diffused to neighboring nodes, which then continues to propagate these embeddings further across the network. Through this layered diffusion process, each node acquires generalized information about distant nodes’ content, enabling the decentralized network to guide queries toward relevant content even when the content is stored multiple hops away.
Our work makes use of these findings and uses diffusion, called ``expansion rounds'', to incrementally improve a node's connections to other nodes, to better fit its embedding.

The query forwarding process of Giatsoglou et al.~\cite{giatsoglou2022graph} is controlled by a time-to-live limit, restricting the number of hops a query can travel to prevent indefinite circulation. When a node receives a query, it calculates a relevance score between the query and the stored node embeddings of neighboring nodes, forwarding the query toward those with the highest scores. Experimental results show that this method effectively finds content within a few hops but encounters reduced accuracy as document density and search radius increase. This makes the method suitable for localized content search in P2P networks, though additional refinements would be needed for scaling to larger document volumes and wider search ranges.
We show that our work requires less hops to retrieve information and operates with larger document volumes.

\section{Design of Semantica}
\label{sec:design}

Semantica places every user in a space whereby their neighbors are interested in the same kind of items that the user is interested in. In order to place the users in this arrangement, we propose an algorithm for constructing a hierarchically-clustered tree in embedding space. 

Each user $u_i \in \mathcal{U}$ is equipped with an identical large language model (LLM) to compute embeddings for their accessed documents $\mathcal{D}_i$. For each document $d_{ij} \in \mathcal{D}_i$, the embedding $\mathbf{D}_{ij}$ is calculated using the LLM:

\begin{equation}
    \mathbf{D}_{ij} = \text{LLM}(d_{ij}).
    \label{eq:document_embedding}
\end{equation}

Subsequently, the embedding of the user, $\mathbf{U}_i$, is determined as the mean of the embeddings of all documents read by that user:

\begin{equation}
    \mathbf{U}_i = \frac{1}{|\mathcal{D}_i|} \sum_{j} \mathbf{D}_{ij}.
    \label{eq:user_embedding}
\end{equation}

Once user embeddings are computed, the construction of the semantic tree begins. The algorithm is formally presented in Algorithm \ref{alg:dynamic_tree_construction}. The algorithm starts by initializing a root node $N_{root}$. Network users $\mathcal{U}$ that are to be added to the tree are processed in a 
loop.  
Initially, the root node is considered a leaf-node which can hold a limited number of users $M$.

\subsection{Tree construction}
\begin{figure}
    \hspace*{-2.7cm}
    \includegraphics[width=400px]{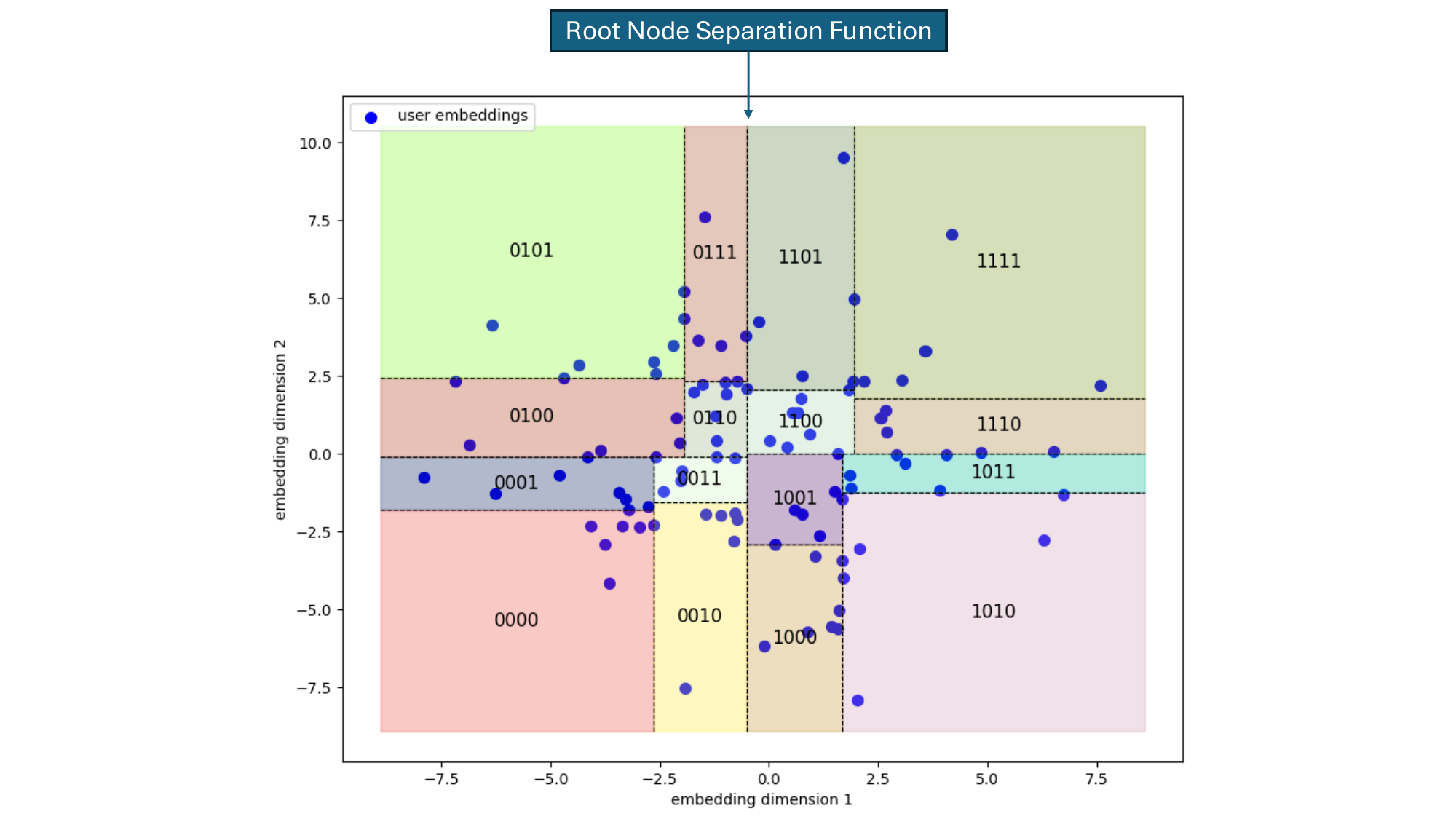} 
    \caption{Example diagram showcasing how the clustering algorithm will construct the tree. The clustering implemented for this figure is a rudimentary one for presentation purposes.}
    \label{fig:example_diagram}
\end{figure}

When the root node is filled with more than that limit $M$, it triggers a K-means clustering process with $k=2$. We have not attempted to vary $k$ during this research project and it was always fixed at $2$. As a result of the clustering, it spawns two child-nodes, each associated with one of the clusters. Users in the current node are then reassigned to one of the child-nodes, corresponding to the cluster they belong to. The root node is no longer a leaf of the tree (since it now has children) and is thus transformed into a split-node. Additionally, we propose a form of ``soft-clustering'' which works by measuring the normed euclidean distance to each of the two centroids in split-nodes. If the difference of the two distances is smaller than a threshold, the user ``clones'' themselves and traverses both paths. We refer to the act of ``cloning'' oneself to mean that the user instead of choosing to go on only one path down the tree, will instead go down both paths. In cloning, the user does not create a new IP for a different position. We use cloning in order to reach a larger number of users with high similarity to us in the tree. In our paper, the tree acts only as a facilitator of efficient information exchange: once the user ends up in a leaf-node, it uses the position to obtain the peers in the leaf-node, and other peers will be able to find the user if they reach the same leaf-node. 

In order to place themselves in the tree, the next user in the loop would navigate the tree by comparing themselves with the centroids of each split-node along the way, cloning themselves if the conditions above are met, until a leaf-node is found which is not yet full.
This method does not require users coming into the network in a particular order, and is able to assign users to leaf-nodes in a dynamic manner resembling peers in real networks entering the system. 


\begin{algorithm}[]
\caption{Dynamic Tree Construction (with Euclidean Dist. \& On-the-fly Cloning)}
\label{alg:dynamic_tree_construction}
\begin{algorithmic}[1]
\REQUIRE Users $\mathcal{U}$, documents $\mathcal{D}$, max users per leaf node $M$, distance threshold $\Delta$
\ENSURE A hierarchically clustered tree $T$ 

\STATE Initialize root node $N_{\mathrm{root}}$, $T = \{N_{\mathrm{root}}\}$
\STATE Shuffle users $\mathcal{U}$ randomly

\FOR{each user $u_i \in \mathcal{U}$}
    \STATE Compute document embeddings $\mathbf{D}_{ij}$ with Equation \ref{eq:document_embedding} for all documents $d_{ij} \in \mathcal{D}_i$
    \STATE Compute user embedding $\mathbf{U}_i$ as per Equation \ref{eq:user_embedding}

    \STATE $n \leftarrow n_{\mathrm{root}}$ \quad \COMMENT{Start tree traversal at the root}
    \WHILE{$n$ is not a leaf node}
        \STATE Let $n_1$ and $n_2$ be the two child nodes of $n$
        \STATE Compute $d_1 = \|\mathbf{U}_i - \mathbf{C}_{n_1}\|_{2}$ \quad \COMMENT{Normed Euclidean distance to centroid of $n_1$}
        \STATE Compute $d_2 = \|\mathbf{U}_i - \mathbf{C}_{n_2}\|_{2}$

        \STATE \COMMENT{Check if $u_i$ is ``close enough'' to both centroids to warrant cloning}
        \IF{$\bigl|\,d_1 - d_2\bigr| < \Delta$}
            \STATE \COMMENT{Clone $u_i$ into the \emph{other} branch as well}
            \STATE \textbf{Clone} $u_i$ into both children $n_1$ and $n_2$
        \ENDIF

        \STATE \COMMENT{Traverse into whichever child centroid is \emph{closer} (i.e., smaller distance)}
        \IF{$d_1 \leq d_2$}
            \STATE $n \leftarrow n_1$
        \ELSE
            \STATE $n \leftarrow n_2$
        \ENDIF
    \ENDWHILE

    \STATE \COMMENT{Now $n$ is a leaf node; assign $u_i$ here}
    \STATE $n.\text{users} \gets n.\text{users} \cup \{u_i\}$

    \IF{$|n.\text{users}| > M$}
        \STATE \COMMENT{Perform $k{=}2$-means clustering on $n.\text{users}$}
        \STATE $\{\mathcal{C}_1, \mathcal{C}_2\} \gets \text{kMeans}(n.\text{users}, k=2)$
        \STATE Let $\mathbf{C}_{n_1}$ and $\mathbf{C}_{n_2}$ be the resulting centroids
        \STATE Create two child nodes $n_1, n_2$ of $n$
        \STATE $n_1.\text{users} \gets \mathcal{C}_1$, \quad $n_2.\text{users} \gets \mathcal{C}_2$
        \STATE Link $n$ in $T$ with its children $n_1$ and $n_2$ 
        \STATE \COMMENT{Leaf node $n$ becomes an internal (split) node}
    \ENDIF

\ENDFOR

\STATE \textbf{Return} the final tree $T$
\end{algorithmic}
\end{algorithm}

For a simplified and intuitive visualization of the hierarchical clustering effect in two dimensions, see Figure \ref{fig:example_diagram}. Each dotted line represents a branching function, separating two branches of a split-node. Each region belongs to a leaf node in the semantic tree and is named according to the branches traveled to reach it. First, a separating line is calculated that splits the tree from the root node: all clusters on its left start with 0 and all clusters on its right start with 1. That represents the first split-node of the tree. On each resulting branch another separating line is calculated, and so on. The users located in the same leaf-node are more likely to hold common documents than users far away in the constructed tree. Thus, the tree-based clustering mechanism presented in this paper allows easier discovery of users who hold similar items. Note that this method returns an approximation of the optimal neighborhoods for users. Such an approximation is a necessary trade-off in a decentralized environment where computation of the optimal neighborhoods (i.e., the set of most similar users to every user) is very costly and thus not practical, due to the communication complexity required to compute it. In later Sections in this paper, we analyze the quality of our approximation empirically and we also analyze the complexity of the proposed computation method.

\subsection{Semantic neighbor discovery}

\begin{algorithm}[]
\caption{Closest Users Maintenance and Expansion Rounds}
\label{alg:closest_users_expansion}
\begin{algorithmic}[1]
\REQUIRE Users $\mathcal{U}$, embeddings $\mathbf{U}_i$ for each user $u_i$, number of closest users for the current user $n_{cu}$
\ENSURE Updated ``closest-users'' and ``known-users'' lists for each user

\STATE \textbf{Initialization:}
\FOR{each user $u_i \in \mathcal{U}$}
    \STATE Gather $n_{cc}$ closest users for each clone of $u_i$ into $u_i.\text{known-users}$
    \STATE Sort $u_i.\text{known-users}$ by cosine similarity to $\mathbf{U}_i$
    \STATE Set $u_i.\text{closest-users}$ as the top $n_{cu}$ users from $u_i.\text{known-users}$
\ENDFOR

\STATE \textbf{Expansion Rounds:}
\FOR{each round $r = 1$ to $r_{\max}$}
    \FOR{each user $u_i \in \mathcal{U}$}
        \STATE Query a random $u_j \in u_i.\text{closest-users}$ for a closer user $u_k \notin u_i.\text{known-users}$
        \IF{$u_k$ is closer to $\mathbf{U}_i$ than the 50th user in $u_i.\text{closest-users}$}
            \STATE Add $u_k$ to $u_i.\text{known-users}$
            \STATE Recompute $u_i.\text{closest-users}$ as the top $n_{cu}$ users from $u_i.\text{known-users}$
        \ENDIF
    \ENDFOR
\ENDFOR

\STATE \textbf{Output:} Return updated $\text{closest-users}$ for all users

\end{algorithmic}
\end{algorithm}

\begin{figure}
    \includegraphics[width=300px]{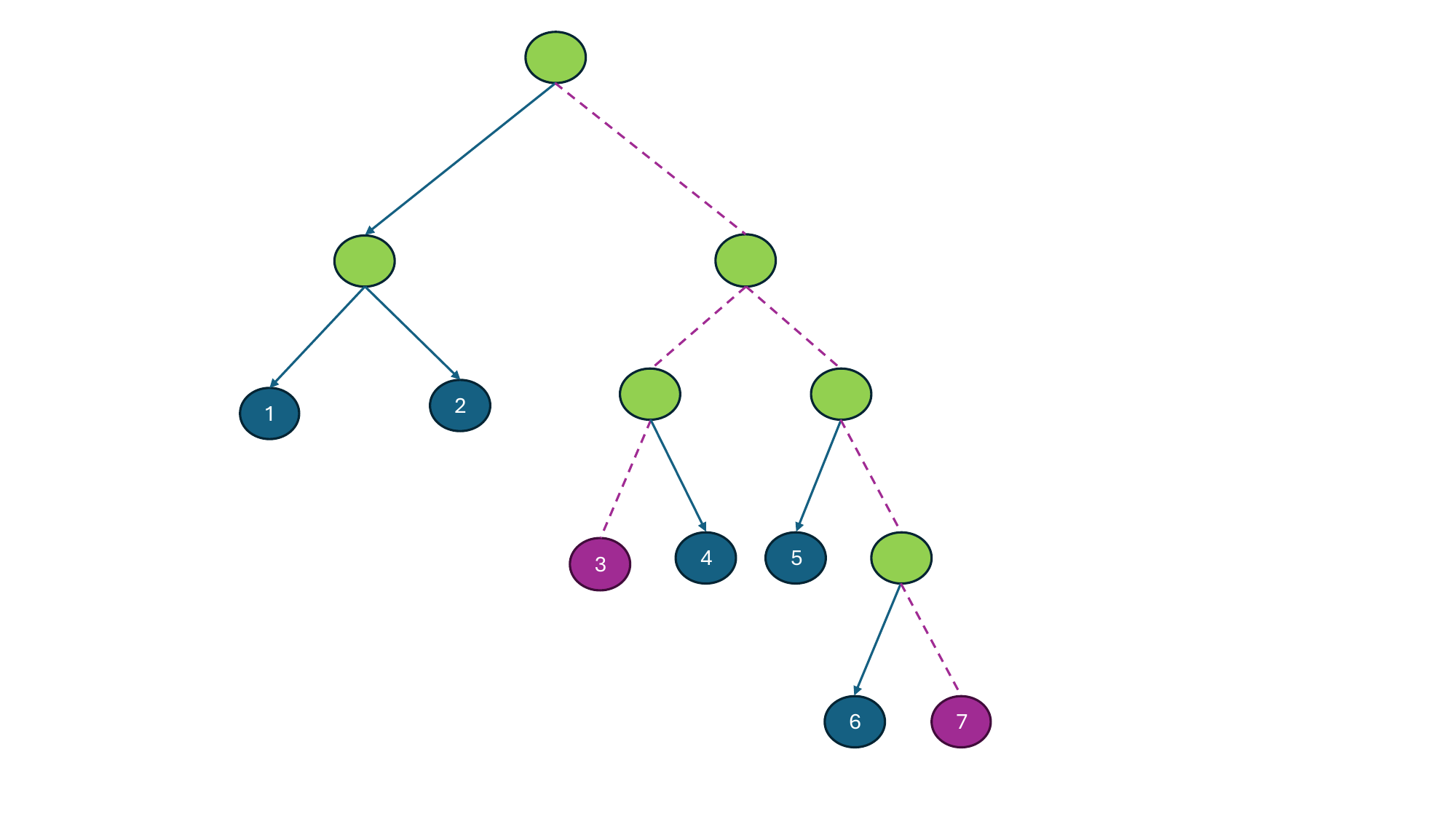} 
    \caption{Tree example diagram. The green nodes represent split-nodes and the blue/purple numbered nodes represent the leaves.}
    \label{fig:tree_diagram}
\end{figure}

Once a user has stabilized into a leaf-node, it acquires the addresses of the $n_{cc}$ nearest neighbors for each of its clones. For a visualization, we refer to Figure \ref{fig:tree_diagram}. We show a hypothetical tree containing a number of split-nodes (colored in green) and leaf-nodes (colored in blue and purple, and numbered). A hypothetical user follows the purple dotted line, clones themselves on the second split-node and ends up in leaf-nodes 3 and 7. Each clone collects the closest $n_{cc}$ other users from their location. If $n_{cc}$ is larger than the number of users placed in the leaf node of a clone, it performs a Breadth-First-Search (BFS) algorithm to find nearby leaf-nodes and collects users placed in them (in a random order) until $n_{cc}$ users are found in the vicinity of each clone. In this case, the clone placed in leaf 3 would first collect users in leaf 4 (as that is the closest leaf-node to leaf 3) and the clone in leaf 7 collects users in leaf 6. If the number of users collected is not large enough, the BFS is performed further, and so on.

Users combine the contacts gathered from all their clones and compute cosine similarity between their own embedding and the embedding of each of their gathered contacts. The top $n_{cu}$ most similar peers are stored in a closest-users list. This list is useful for limiting the number of contacts queried, reducing strain on the network. The most similar peers are also the ones most likely to have the relevant documents.

While the tree structure helps cluster similar users into the same leaf nodes, recall that it is an approximation. Due to the nature of hierarchical clustering, some peers who are highly similar may end up in different leaf nodes. To address this, peers periodically ask a randomly chosen user from their closest-users list if they know anyone who is more similar to them than the $n_{cu}$'th closest (i.e., the least close) user in their list. If so, they are added to the peer’s known-users list, improving the accuracy of their closest-users selection and overall network connectivity. This procedure follows the pseudo-code in Algorithm \ref{alg:closest_users_expansion}.

Unlike tree construction, which uses Euclidean distance to cluster users, expansion rounds rely on cosine similarity to refine user connections. This ensures that search results align with semantic relationships rather than just spatial proximity.

This process is simultaneous (asynchronous) for all peers and we refer to it as one expansion-round.

\subsection{Query Model}
Once the network is established and each user has constructed their closest-users list, queries can be performed to search for desired documents. The proposed query mechanism, termed \emph{chain-hopping}, dynamically adapts the query path based on semantic similarity. This approach aims to minimize network load while maintaining high accuracy.

In chain-hop querying, a user initiates a query by contacting itself. If the desired document is not found in the user's local storage, the query is forwarded to another user from the user's closest-users list who is \emph{closest} to the query embedding (in terms of cosine similarity). If the document is not found in this other user's storage too, the process repeats iteratively, with each user forwarding the query to the next most semantically relevant user, until the desired document is found or a predefined maximum hop limit is reached. We have tested other querying methods but have found that this one retrieves most documents regardless of the graph in which it is implemented, so we only present the result for this querying-method.
This method ensures that the query traverses the network in a targeted manner, focusing only on users likely to hold the desired document. The adaptive nature of chain-hopping allows it to efficiently discover documents even in a highly distributed and dynamically evolving network.

\begin{itemize}
    \item \textbf{Network Overhead.} Each hop traversal involves a single message being sent from the current user to the next closest user in the query chain. For a query traversing up to $\ell$ hops, the network overhead is $\mathcal{O}(\ell)$ messages.
    \item \textbf{Local Computation Cost.} At each hop, the current user calculates the cosine similarity between the query embedding and the embeddings of all users in their closest-users list. For a list of size $k$, this incurs a local cost of $\mathcal{O}(k)$ per hop. Across $\ell$ hops, the total local cost is $\mathcal{O}(\ell \cdot k)$.
    \item \textbf{Advantages.} The chain-hopping method is both resource-efficient and highly adaptable, as it prioritizes semantically relevant paths through the network without flooding unnecessary nodes.
\end{itemize}

\begin{algorithm}[]
\caption{Chain-Hop Query Processing}
\label{alg:chain_hop_query}
\begin{algorithmic}[1]
\REQUIRE Query embedding $\mathbf{Q}$, max hop limit $\ell$
\ENSURE Matching documents $R$

\STATE Initialize current user $u = u_q$ (query initiator)
\STATE Initialize hop count $h = 0$

\WHILE{$h < \ell$ \textbf{and} required document not found}
    \STATE Query $u$ for matching documents
    \IF{match found}
        \STATE \textbf{Return} matching documents $R$
    \ENDIF
    \STATE Find $u_{\text{closest}} = \arg\max_{v \in u.\text{closest-users}} \text{Similarity}(\mathbf{Q}, \mathbf{U}_v)$
    \STATE Update $u = u_{\text{closest}}$
    \STATE Increment hop count: $h \gets h + 1$
\ENDWHILE

\STATE \textbf{Output:} Return $R$ (empty if no match found within $\ell$ hops)

\end{algorithmic}
\end{algorithm}

\section{Implementation and Performance Analysis}
\label{sec:impperformance}

This section outlines the methods used to evaluate the proposed system. The experiments focus on measuring the success of the decentralized semantic tree structure in facilitating efficient and accurate document retrieval. Key evaluation metrics include query accuracy within a certain number of queries and the effectiveness of clustering mechanisms. The evaluation methods are designed to test both the quality of the tree structure and the performance of different query mechanisms under varying conditions.

\subsection{Experimental Setup}
Our experimental setup requires data preparation, implementation of embeddings, parameterization of our tree and of new user discovery.
We now define our methods and values to achieve them.

\begin{enumerate}
    \item \textbf{Data preparation and user embeddings:}
    \begin{itemize}
        \item Records where a user accessed the same document multiple times were filtered out and only the first access by timestamp was chosen, corresponding to the query which first led the user to the target website. Users which had less than 30 unique documents in their search history were filtered out.
        \item Each user’s documents were randomly split into a ``test'' set (10 documents) and a ``training'' set (the remaining documents). The assignment of each document to either the train or test set was entirely random. Since documents in this study correspond to URL links, different sub-pages of the same domain were treated as distinct documents. As a result, sub-pages from the same domain could appear in both the train and test sets. However, we do not consider this to be a form of data leakage, as different URLs have unique titles---the content scraped during the construction of AOL4PS---which results in distinct embeddings.\footnote{In future work, other splitting criteria (e.g. based on timestamps) may also be considered. Random splitting is common in related datasets such as the MovieLens dataset.}
        \item User embeddings are computed as the mean embedding of their training document titles.
    \end{itemize}
    \item \textbf{Tree construction:}
    \begin{itemize}
        \item Users are added sequentially to a root node, with nodes splitting when they exceed a capacity of \textbf{L} = 50 users.
        \item Users near the cluster boundary (difference between the distances to both centroids $< \Delta$) are cloned into both child nodes during splits.
        \item Each user computes a ``known-users'' list based on proximity to other users in the same or nearby leaf nodes.
        \item From the ``known-users'' list, each user generates a "closest-users" list containing the $n_{cu}$ most similar users.
    \end{itemize}
    \item \textbf{Known-users expansion:}
    \begin{itemize}
        \item Users iteratively refine their ``closest-users'' list by querying random neighbors in their ``known-users'' list to identify closer peers based on cosine similarity.
        \item Expansion rounds continue until a pre-specified number of interactions is reached or the ``closest-users'' list stabilizes.
    \end{itemize}
\end{enumerate}

\subsection{Experiment designs}
\label{sec:experiments}

We design three experiments to evaluate the performance of \algorithmname{}.
The first experiment measures how fast connected users are identified as semantically-similar.
The second experiment measures the communication required to retrieve documents.
The final experiment targets the efficacy of \algorithmname{}, in respect of the number of semantically-similar users that are connected.

\textbf{Experiment 1: Closest-User Identification with a Limited Number of Interactions.}
This experiment evaluates the ability of \algorithmname{} to identify the addresses of users who are closest in embedding space to a given user, as measured by cosine similarity. To establish a baseline, we calculate the cosine similarity between each user and every other user in the dataset, generating a global list of the 50 closest peers for every user based on these cosine similarity values. This global list represents the ground truth for evaluation.

After constructing the tree using \algorithmname{}, we compute the \textit{closest-users} set for each user by relying solely on the local neighbor information and interactions facilitated by the tree structure (as per Algorithms \ref{alg:dynamic_tree_construction} and \ref{alg:closest_users_expansion}). The accuracy of this set is then assessed by calculating the intersection between the \textit{closest-users} set derived from \algorithmname{} and the ground truth list of the 50 closest peers. The absolute size of this intersection, serves as a measure of \textit{absolute-recall-out-of-50} which will be referred to as \textit{recall} in the rest of the text.

To explore the impact of different parameters, we compare the recall obtained for various values of $\Delta$. Additionally, we contrast these results with the recall achieved when performing expansion rounds on a Barabási–Albert graph. The \textbf{m} parameter of the Barabási–Albert graph, which controls the average degree of the graph, is adjusted to match the average graph size obtained using \algorithmname{}. This experiment showcases the speed of identification of the specific users who are closest in embedding space to a peer when using \algorithmname{} vs performing this identification randomly. 

\textbf{Experiment 2: Document Retrieval Mechanisms.}
The effectiveness of \algorithmname{} is assessed through a query mechanism inspired by the method presented in the work of Giatsoglou et al.\cite{giatsoglou2022graph}, which evaluates the efficiency and accuracy of document retrieval. The retrieval process begins with a query initiated by a user seeking a document. The query is first sent to the user whose embedding is closest to the query embedding of the requester. 

If the queried user does not have a matching document in their local document set, they proceed to evaluate their \textit{known-users} list. Each neighbor in the list is compared against the query embedding using cosine similarity, and the query is forwarded to the neighbor with the highest similarity score. This forwarding process continues iteratively, up to a predefined maximum of \textbf{$max\_hops$} hops.

After all users in the network have sent the queries from their respective test sets, the performance of the system is measured as the percentage of queries successfully resolved within the specified hop limit. This metric provides a clear indication of the system’s ability to locate and retrieve documents efficiently under the constraints of the decentralized network.

\textbf{Experiment 3: Smallest Distance to a Target Document.}
The effectiveness of \algorithmname{} can also be evaluated by measuring the minimum number of hops required to locate a specific document from the position of a randomly selected user. Unlike the query-based evaluation in the previous experiment, this analysis does not rely on targeted querying of the closest user to the query embedding. Instead, it considers the absolute distance, in terms of hops, to any user who possesses the required document.

An ideal algorithm would organize peers in such a way that the majority of documents are accessible within a small number of hops for most users, while maintaining a constant average number of connections per peer. To assess this, we construct a directed graph using \algorithmname{} and compare it to an equivalent Barabási-Albert graph. The \textbf{m} parameter in the Barabási-Albert graph, representing the number of edges added for each new node, is selected to match the average degree of the graph created by \algorithmname{}.

This comparison provides a metric for the efficacy of \algorithmname{} in respect of minimizing hop distances while preserving the structural efficiency of the network.



\section{Results and discussion}
\label{sec:resultsdiscussion}
This section presents the results of the three experiments described in Section \ref{sec:experiments}.

\subsection{Closest User Recall}

The purpose of this experiment is to determine the efficiency of discovery of most similar peers, focused specifically in a converged network.

\begin{figure}
    \centering
    \begin{subfigure}[b]{0.45\textwidth} 
        \includegraphics[width=\textwidth]{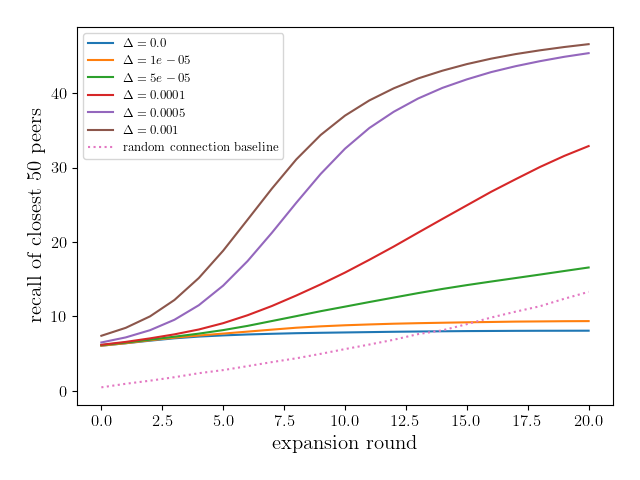}
        \label{fig:recall/nbr_known_users_filterFalse}
    \end{subfigure}
    \hfill
    \caption{Effect of expansion rounds on closest-neighbor recall for different $\Delta$ values.}
    \label{fig:cc-perexpansion}
\end{figure}

We refer to the number of closest-peers successfully found as ``closest user recall''. In Figure \ref{fig:cc-perexpansion} we show the result of this analysis. We can see that constructing the tree without expansion rounds does not yield a high mean recall per user, though increasing $\Delta$ does help. However, when $\Delta$ is larger than zero, asking people in our closest neighborhood increases the mean recall substantially, to the point where if we undertake 20 rounds of known-users expansion we can find almost all of the required users (more than 40 for all $\Delta$'s).

But under normal circumstances we can also get high recall by just asking peers randomly, without taking into consideration the tree structure.

To investigate this, we implemented a simulation where each peer had the same number of known-users as they did at the start of the $\Delta=0.001$ experiment (before the expansion rounds), but the list of known-users is randomized for each peer. Then, we followed the expansion-rounds algorithm the same way as when constructing Semantica. We chose $\Delta=0.001$ as the representative parameter for the size of the graph because it was the experiment with the largest number of known-users before any expansion round was performed. This means that by obtaining a higher recall than the baseline with any of the $\Delta$ values shown in Figure \ref{fig:cc-perexpansion}, the tree algorithm increases the chance of finding the closest peers in cosine-similarity space.

Figure \ref{fig:cc-perexpansion} highlights the significant advantage provided by the tree structure, even before any expansion rounds take place. Semantica outperforms the baseline by an order of magnitude at initialization: the random method achieves a recall of approximately 0.4, whereas all Semantica configurations exceed 5.
As expansion rounds progress, Semantica continues to maintain a substantial lead. By round 10, the random baseline reaches a recall of about 5, while Semantica achieves a recall of approximately 35 for $\Delta = 0.001$ and 30 for $\Delta = 0.0005$. However, for $\Delta = 0$, the recall does not increase significantly with additional expansion rounds. This limitation arises because, with a very limited or non-functional cloning mechanism, users within the same leaf-node share identical neighbors. Consequently, querying neighbors reveals largely redundant information, limiting the effectiveness of expansion rounds.
The benefit of expansion rounds becomes apparent when cloning and moderate $\Delta$ values are utilized. These mechanisms introduce diversity in the network by ensuring users have some disjoint contacts, enabling meaningful information exchange and thus improving recall performance.
Without this function, expansion rounds would not work when combined with the tree as it is presented in this paper.

\begin{figure}
    \centering
    \includegraphics[width=0.45\textwidth]{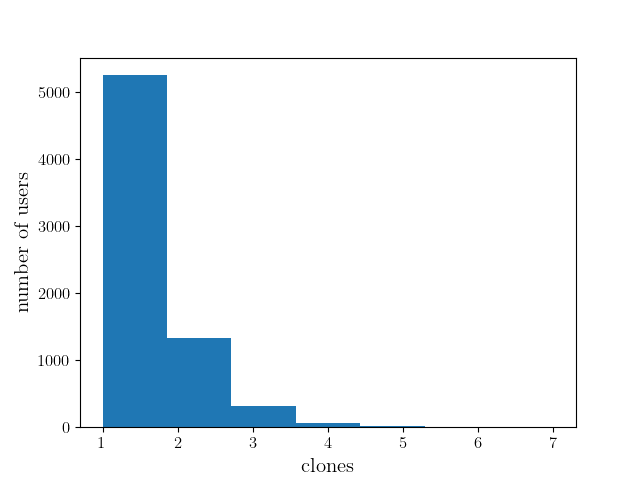}
    \caption{Number of clones per user for $\Delta = 0.001$.}
    \label{fig:cc/clones_per_user}
\end{figure}

Figure \ref{fig:cc/clones_per_user} illustrates the distribution of clones per user at $\Delta = 0.001$, the upper threshold identified in Figure \ref{fig:cc-perexpansion}. The distribution reveals a clear decreasing exponential trend, with approximately 5100 of the 6980 users remaining unaffected by the cloning mechanism. This result suggests that $\Delta$ can be set high enough to significantly improve the recall of similar users without causing a substantial increase in system complexity. The modest number of clones and their limited impact on the majority of users underscores the efficiency of the chosen configuration.

Table \ref{tab:nbr_clones_per_user_for_different_delta} provides a detailed breakdown of the expected number of clones per user across various $\Delta$ values. The results indicate that at approximately $\Delta = 0.001$, the number of clones per user begins to increase more rapidly, signaling a notable rise in the computational burden imposed by soft clustering. When $\Delta$ reaches 0.005, the expected number of clones per user exceeds 7, potentially amplifying the complexity for that user significantly. While these findings are specific to the AOL4PS dataset, the exponential growth in clones as $\Delta$ increases is likely to occur in other datasets as well, though the threshold at which this growth becomes significant may vary depending on the data characteristics.

\begin{table}[h!]
    \centering
    \caption{Number of clones per user for different $\Delta$}
    \label{tab:nbr_clones_per_user_for_different_delta}
    \begin{tabular}{lrrrr}
        \toprule
        \textbf{Delta} & \textbf{Median} & \textbf{Mean} & \textbf{Std} \\
        \midrule
         0 & 1 & 1.0000 & 0.00 \\
         1e-6 & 1 & 1.0001 & 0.01 \\
         5e-6 & 1 & 1.0012 & 0.03 \\
         1e-5 & 1 & 1.0027 & 0.05 \\
         5e-5 & 1 & 1.0127 & 0.11 \\
         1e-4 & 1 & 1.0259 & 0.16 \\
         5e-4 & 1 & 1.1489 & 0.41 \\
         1e-3 & 1 & 1.3224 & 0.64 \\
         3e-3 & 2 & 2.7560 & 2.25 \\
         5e-3 & 5 & 7.4497 & 8.08 \\
        \bottomrule
    \end{tabular}
\end{table}

\subsection{Document retrieval accuracy with query chain-hopping}
This experiment was meant to determine the effectiveness of query hopping while utilizing Semantica as opposed to random query hopping in a random graph of similar structure (same mean degree of nodes). We also compared the accuracy of the graph-diffusion algorithm proposed by Giatsoglou et al.~\cite{giatsoglou2022graph} in the same graph for different values of teleportation-chance. The teleportation-chance determines the level of diffusion performed in the graph.

\begin{figure}
    \centering
    \begin{subfigure}[b]{0.45\textwidth} 
        \includegraphics[width=\textwidth]{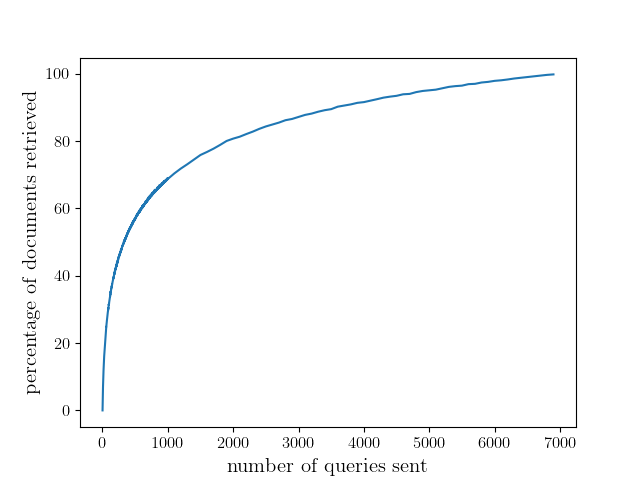}
        \caption{Random Query Baseline}
        \label{fig:Random_query_baseline}
    \end{subfigure}
    \hfill 
    \hspace{1cm}
    \begin{subfigure}[b]{0.45\textwidth} 
        \includegraphics[width=\textwidth]{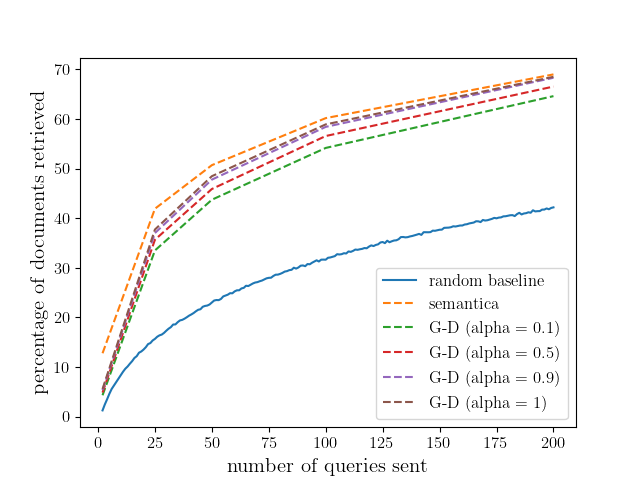}
        \caption{Comparison of retrieval rate at 50 queries sent to the network}
        \label{fig:Accuracy_comparison}
    \end{subfigure}
    \caption{Accuracy Comparison. }
    \label{fig:cc/accuracy_comparison}
\end{figure}

The results for the chain-hop method versus the graph-diffusion method, for various alpha values, are shown in Figure \ref{fig:cc/accuracy_comparison}. Figure \ref{fig:Random_query_baseline} shows the accuracy at different number of queries sent to the network if queries are sent to random peers. So, assuming everybody knows everybody else, and each person chooses to send $n_q$ queries in the network and chooses the peers randomly, we get the accuracy curve shown in Figure \ref{fig:Random_query_baseline}.
Figure \ref{fig:Accuracy_comparison} shows the accuracy for the random-baseline, Semantica with $\Delta$ = 0.003 and 10 expansion rounds, and graph-diffusion for between 2 and 600 queries. Graph-diffusion is implemented in a Barabási-Albert graph with \textbf{m} = 104. The m=104 was chosen because it creates a graph with the same amount of connections as the mean value of ``Known-Users'' for Semantica with the parameters specified above (i.e. approximately 208).
From the figure we can observe that:
\begin{itemize}
    \item Graph-diffusion with alpha = 1.0 performs better than alpha = 0.9, which is better than alpha = 0.5, which itself performs better than alpha = 0.1. This indeed follows the results of the authors of the paper showing that higher levels of alpha (i.e. lower levels of diffusion) generally lead to better results. Including any diffusion in the system degrades the retrieval accuracy. So the actual algorithm to beat is the simple query chain hop with a graph constructed randomly.
    \item  When we consider a maximum of two-hops per query, Semantica retrieves 12.75\% of required documents while any of the other methods achieve less than 6\%. The larger the number of queries, the smaller the magnitude of the difference. This is because as the number of hops available increases, the initial position of the node matters less and less. The initial position of the node determines an increase of the retrieval accuracy because the neighborhood constructed through Semantica increases the chances that nearby users have our required document. If they do not, after the hop takes a few wrong turns, we arrive in a situation where the query is now in another neighborhood altogether, and from here all benefits of using the semantic tree disappear. 
\end{itemize}

\subsection{Semantica tree versus diffusion graph trade-offs} 
\begin{figure}
    \centering
    \begin{subfigure}[b]{0.45\textwidth} 
        \includegraphics[width=\textwidth]{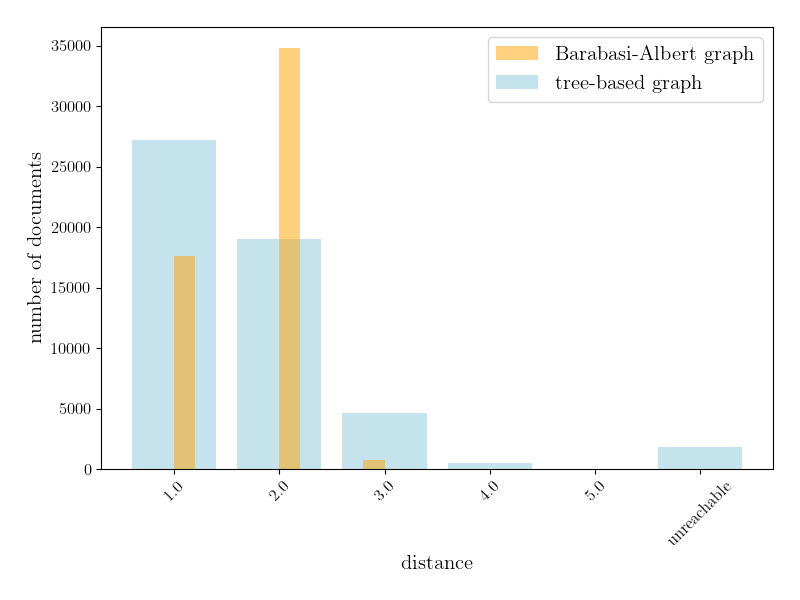}
    \end{subfigure}
    \hfill
    \caption{Minimum distance to the correct document for tree with $\Delta = 0.001$ and 10 expansion-rounds, and a random Barabási-Albert graph with $m = 86$.}
    \label{fig:min_golden_distance}
\end{figure}

To evaluate the performance of the \algorithmname{} algorithm in a scenario where the graph-diffusion algorithm is enhanced, we constructed Figure \ref{fig:min_golden_distance}. In this figure, we first use the \algorithmname{} algorithm to build a directed graph based on the known-users lists of each peer in a tree with $\Delta=0.001$ and 10 expansion rounds. We then generate a Barabási-Albert graph with an m-value chosen to match the mean degree of the known-users graph produced by the tree algorithm.

With these graphs constructed, we calculate the minimum hop distance to a node which has the required document. Figure \ref{fig:min_golden_distance} presents the Barabási-Albert graph in yellow and the Tree-based graph in light blue. The tree algorithm, as expected knowing the results presented so far, shifts many documents from a distance larger than one to a distance equal to one. But there is also another effect which shifts the documents at a distance of 2 to a distance of 3 (with few at a distance of 4 and 5), and also makes some documents unreachable (since some sub-graphs are unconnected to the rest of the network). 

We saw in Figure \ref{fig:cc/accuracy_comparison} that the graph created with Semantica yielded better results up to 200 hops per query. It is then to be expected that as the number of hops increases heavily (to the point where we reach a significant section of the entire population), that the Barabási-Albert based graph would start overtaking the Semantica-based graph simply because some of the documents are unreachable in it.

The lengthening of the required hop-distance effect is due to the fact that peers cluster together and have each other in their known-users list. This means that the peers in a group are less likely to hold connections outside of the group. If a document which is not part of the interests of the group is required, then the distance to the closest peer who has that document could very well go up. While the tree method improves the accuracy overall, if we would like to consider complex chain-hopping techniques, it would be best to take into consideration a combination of meeting users with the tree method and with a random meeting procedure. This would connect disparate groups and ensure that users could get easier through the network.

\section{Algorithm Complexity}
\label{sec:complexity}

This section analyzes the complexity of building and maintaining the proposed semantic tree 
(Algorithm \ref{alg:dynamic_tree_construction}) 
and the complexity of processing queries 
(Algorithm \ref{alg:closest_users_expansion} 
and our querying methods). We consider both the \emph{average} scenario, in which user embeddings are reasonably well-distributed (yielding a near-balanced tree), and a \emph{worst-case} scenario, in which the data distribution is extremely skewed (e.g., most users cluster in a single region of the embedding space). Throughout our analysis, embedding size is a constant factor (768 dimensions) and not shown explicitly.

\subsection{Complexity of Tree Construction}

Our algorithm is based on four distinct functionalities: insertion of new users, leaf-splitting, user cloning, and expansion-rounds.
We now discuss their individual time complexity.

\textbf{Insertion of Users into the Tree.}
Each user is inserted by starting at the root and traversing down to a leaf node, comparing the user’s embedding with the child-node centroids at each split-node.
The average case to be investigated is the situation where the tree is \textit{near-balanced}. If splits are fairly balanced, the tree height is $\mathcal{O}(\log N)$, where $N$ is the number of users. Inserting one user costs $\mathcal{O}(\log N)$ in traversal, so inserting all $N$ users totals $\mathcal{O}(N \log N)$. Network-wide, each insertion takes $\log N$ hops from root to leaf.
    
In some situations, the tree could potentially be \textit{unbalanced}. In order for this situation to occur, we need a large number of users at each level of the tree to be significantly closer together and few users to be outside of the main cluster. It is conceivable that a large number of leaf-nodes would contain exactly one person. This would require that, assuming we start with 50 users, the 51'st user is very far away. This would lead to the centroid having the same coordinates as the user itself, thus leading to a leaf-node of 1 person. In order for this situation to occur on a large scale, each split must obey this sort of structure. But this structure is itself assuming that every split would have a cluster of almost 50 people and another cluster of 1 or 2. It also assumes that no new users would go down that path either. This specific structure is very unlikely to occur in real situations at multiple depths of the tree, thus the complexity can be assumed to be in the vast majority of situations $O(logN)$. In the very unlikely scenario where this structure of user embeddings occurs, the tree would be $\mathcal{O}(N)$ in height, costing $\mathcal{O}(N)$, leading to $\mathcal{O}(N^2)$ overall.

\begin{figure}
    \centering
    \begin{subfigure}[b]{0.45\textwidth}
        \includegraphics[width=\textwidth]{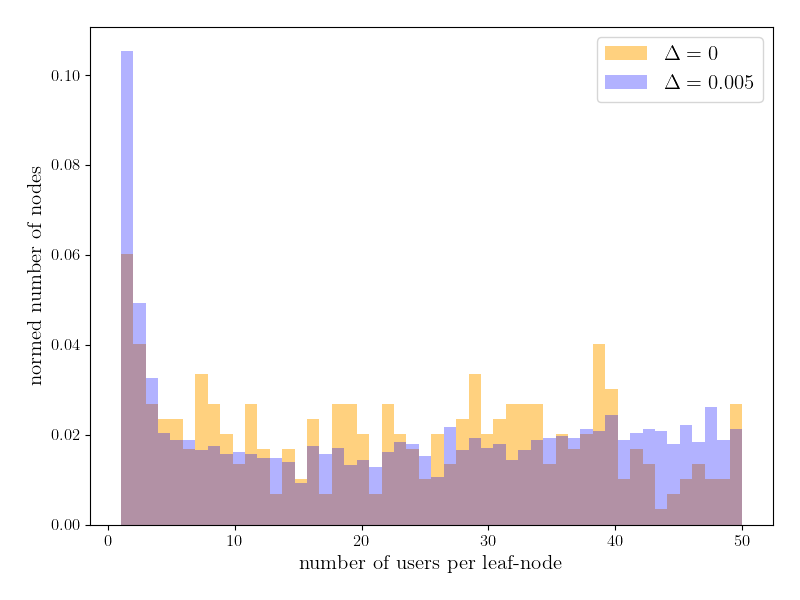}
        \label{fig:recall/fig8_normed_users_per_leaf}
    \end{subfigure}
    \caption{Normalized number of users per leaf node for $\Delta \in \{0, 0.005\}$ (\textit{filtered} dataset). 
    A larger $\Delta$ increases the fraction of leaves with very few users leading to a more skewed tree.}
    \label{fig:normed_usersperleaf}
\end{figure}

To illustrate potential skew or uniformity in leaf sizes, Figure \ref{fig:normed_usersperleaf} shows a histogram of the normalized number of users per leaf node for two different $\Delta$ values. Larger $\Delta$ tends to create more leaves with very few users, indicating some unbalanced splits. Nevertheless, one can see that the majority of leafs do hold about the users following a uniform distribution, even for very large $\Delta$. 

\textbf{Complexity of Leaf Splits Clustering.}
A leaf that exceeds $M$ users triggers $k{=}2$ means clustering on its user embeddings to produce two child nodes. Each split reassigns up to $M$ users. $k{=}2$ has cost $\mathcal{O}(M \cdot I)$, where $I$ is the number of iterations in the k-means clustering ($I$ is often small for moderate $M$). Because leaf nodes only split after exceeding $M$ users, the total number of splits is generally $\mathcal{O}(N / M)$. Summed over all splits, clustering cost is $\mathcal{O}(N \cdot I)$, typically dominated by the $\mathcal{O}(N \log N)$ insertion cost if $I$ is small.

\textbf{Complexity Incurred by User Cloning.}
If a user is within threshold $\Delta$ of multiple child centroids, it clones into both subtrees. Let $c_{\text{avg}}$ be the average number of clones per user:
\begin{itemize}
    \item With moderate $\Delta$, $c_{\text{avg}}$ would be a small constant, yielding a message complexity of $\mathcal{O}(c_{\text{avg}} \, N \log N) \approx \mathcal{O}(N \log N)$ total. Computational complexity would also be multiplied by $c_{avg}$ (as each clone would need to compare itself with downstream centroids) and treated in exactly the same manner.
    \item If $\Delta$ is huge and the space is nearly uniform, cloning can be excessive, but practical implementations such as ours cap $\Delta$ or would detect duplicates to prevent runaway splitting.
\end{itemize}

\textbf{Complexity of Expansion Rounds.}
After building the tree, each user has an initial \textit{known-users} list (obtained through leaf peers and BFS to nearby nodes). In each \emph{expansion round}, every user queries one of its known-users for any closer peers, updating its closest-users list.
\begin{itemize}
    \item Each round involves $N$ messages total (one per user).
    \item Over $r_{\max}$ rounds, that is $\mathcal{O}(r_{\max} \, N)$ messages. Local updates per user are small (inserting/removing in a $K$-size priority queue). Computing similarity scores between the requesting person and all $k$ neighbors incurs $O(k)$ cost for each round so $O(r_{max}k)$.
\end{itemize}
All in all, tree construction plus expansions is $\mathcal{O}(N \log N + r_{\max} \, N)$ in messages and $O(N \log N + N \, r_{max} \, k_{mean})$ in computation requirements where $k_{mean}$ is the mean number of neighbors for all users. In terms of network messages, we can expect that under a large $N$, the $N \log N$ term will dominate. However, in terms of computation requirements, assuming each peer has a large number of known-users and the system performs a sizable $r_{max}$ expansion rounds, the $N \, r_{max} \, k_{mean}$ term may dominate. 

The alternative to our expansion method would be to skip the computation of similarity and just share every neighbor with every asking person. However, this will exponentially increase the number of neighbors of every participating user, thereby overwhelming the network in terms of messages.

\subsection{Query Complexity}

Once the overlay is established, queries are processed using a chain-hop method, where the query traverses the network hop-by-hop to locate the most relevant document. Each query originating from a user is forwarded iteratively through the network based on semantic similarity. At each hop, the current user evaluates if there is any document which would be a satisfactory result to the query. Local keyword-search would benefit from indexing but we assume some kind of semantic-based search. In this case we can expect a complexity of $ \mathcal{O}(d)$ where $d$ is the expected number of documents of a user. If there is no relevant document found, the current user computes the cosine similarity between the query embedding and the embeddings of all its known neighbors. Then the current user forwards the query to the neighbor most similar to it. This process continues until the desired document is located or a maximum number of hops, \(\ell\), is reached.
In summary, the complexity of this query method is defined by the following two points.

\begin{itemize}
\item \textbf{Network Overhead.} Each query requires at most \(\ell\) messages, as the query is forwarded through up to \(\ell\) hops in the network. For a total of \(Q\) queries, the overall network overhead is \(\mathcal{O}(Q \cdot \ell)\).

\item \textbf{Local Computation.} At each hop, the current user evaluates the cosine similarity between the query embedding and the embeddings of all \(k\) neighbors. This incurs a local cost of \(\mathcal{O}(d  +  k)\) per hop. Over \(\ell\) hops, the total local computation per query is \(\mathcal{O}(\ell \cdot (k + d))\) .
\end{itemize}

In summary, the chain-hop method scales efficiently with the number of queries and the size of the network, as the overhead is linear in both \(Q\) and \(\ell\). Furthermore, its localized computations ensure that individual users only process information about their immediate neighbors, keeping the method practical even for large-scale decentralized networks.

\subsection{Summary and Remarks}
\label{subsec:complexity_summary}

Overall, our system achieves scalable performance under realistic embedding distributions and moderate cloning thresholds. We derive five complexity results.

In a typical near-balanced scenario, the tree height is $\mathcal{O}(\log N)$, and inserting $N$ users costs $\mathcal{O}(N \log N)$. In a rare, highly skewed distribution, the tree can degenerate to height $\mathcal{O}(N)$, yielding $\mathcal{O}(N^2)$ total insertion cost. Clustering leaves of size above $M$ via $k$-means further mitigates extreme imbalance in real-world conditions.

After the tree is built, each user refines its local view by running $r_{\max}$ expansion rounds. Each round involves $\mathcal{O}(N)$ messages overall, and a local computation cost of $\mathcal{O}(k)$ per user per round to compare embeddings. Hence, expansions contribute up to $\mathcal{O}(r_{\max} \, N)$ in messages and $\mathcal{O}(r_{\max} \, k \, N)$ in local cost. This remains manageable so long as $r_{\max}$ and $k$ remain small compared to $N$.

\label{paragraph:cloning_complexity}
If a user lies near multiple centroids, it “clones” into more than one subtree. As long as the average number of clones, $c_{\text{avg}}$, is bounded by a small constant (e.g.\ $\leq 10$), the insertion cost remains $\mathcal{O}(N \log N)$. Unbounded $\Delta$ in a nearly uniform space could inflate $c_{\text{avg}}$, but practical caps on $\Delta$ or duplicate detection avert blow-ups.

For queries, the network overhead per query is $\mathcal{O}(\ell)$ messages, and each user’s local cost at each hop is $\mathcal{O}(k + d)$, covering similarity checks with $k$ neighbors plus a $d$ local document lookup. Over $\ell$ hops, this yields $\mathcal{O}(\ell (k + d))$ local cost per query, and $\mathcal{O}(Q \cdot \ell)$ total messages for $Q$ queries. Such localized lookups keep the method efficient, even at large scale.

Under moderate $\Delta$ and typical data, building and expanding the semantic tree runs in $\mathcal{O}(N \log N)$ (plus $\mathcal{O}(r_{\max} N)$ for expansions), while chain-hop queries incur only $\mathcal{O}(\ell)$ messages and $\mathcal{O}(\ell (k + d))$ local computation per query. In extreme skew, insertion can degrade to $\mathcal{O}(N^2)$, although in such a uniform embedding space, simpler random or flooding-based mechanisms might suffice.

\section{Deployment Considerations}
\label{sec:decentralized_impl}

The simulations in 
Algorithms \ref{alg:dynamic_tree_construction}, 
\ref{alg:closest_users_expansion}, 
and \ref{alg:chain_hop_query}
demonstrate our approach under controlled conditions. Having presented positive results from Section \ref{sec:resultsdiscussion}, we now outline how the same concepts can be deployed in a \emph{fully decentralized} environment. In such a setting, no global coordinator or centralized server exists, and each peer operates autonomously. The key assumption dropped is that peers do not have universal knowledge of the network (e.g., centroids); they only discover relevant information through local interactions.

\textbf{Custodians and Node Splits.}
In a fully decentralized version, peers obtain centroid information for split-nodes through a type of peer we call \emph{custodian}.
When a leaf node splits into a \emph{split-node}, one peer is selected (randomly or by consensus) to act as the \emph{custodian} of that split-node. The custodian stores the newly created centroids and maps each centroid to the network address (e.g., IP or overlay address) of the corresponding child node. It then notifies the custodian of its parent node (if one exists), thereby maintaining a chain of custodians from the root down to all leaves. This mechanism ensures that future peers traversing the tree can locate the relevant centroids at each level without requiring a global index. A custodian thus holds all centroids generated \emph{within} its split-node, but not necessarily those of unrelated splits occurring elsewhere in the tree.

\textbf{Initialization and the Root Peer.}
A single peer, referred to as the \emph{root peer}, begins the construction of the tree by creating an initial leaf node. Other peers discover the root peer through out-of-band mechanisms such as a bootstrap server, a public DHT, or a known reference address. Although this step relies on a well-known root, the root peer does not serve as a permanent coordinator or a single point of failure. Multiple root peers can independently host the same tree. Additionally, multiple trees can coexist, each dedicated to a particular topic or language. Peers may opt to join any number of these trees, allowing for diverse overlays within the same network.

\textbf{Joining the Tree.}
When a new peer joins, it first requests the \emph{centroid} (or centroids) maintained at the entry point of the tree. If the root node remains a leaf, the peer can be admitted directly. Otherwise, the peer evaluates its \emph{normed Euclidean distance} to each child centroid of the current split-node. It then contacts the custodian responsible for the node that best matches its embedding (and can contact more than one if the distance difference is below $\Delta$). This process repeats until the peer (and all of its clones) reaches a genuine leaf-node and is inserted there. If the leaf-node’s user count exceeds $M$, the peers at that leaf run $k$-means clustering (based on local consensus) to produce two new centroids and split the leaf into two child nodes. All members are reassigned according to their distance to the new centroids and a custodian for the new split-node is chosen.

\textbf{Neighbor Expansion.}
After a peer has joined the tree, it periodically executes the neighbor-expansion process (\emph{cf.} Algorithm \ref{alg:closest_users_expansion}) to refine its local view of the network. During each expansion round, the peer queries one of its known neighbors for potential peers that lie closer in embedding space, thereby updating its \texttt{closest-users} list. This step aligns with the decentralized spirit of the algorithm: no global knowledge is assumed, and each peer maintains a limited neighbor set discovered incrementally. 

\textbf{Considerations and Potential Extensions.}
The decentralized approach outlined here is flexible but also faces real-world challenges. If a custodian departs (voluntarily or otherwise), its centroid data becomes unavailable. A backup custodian election mechanism, such as assigning a secondary custodian, could mitigate this issue. Likewise, although multiple root peers permit diverse trees to exist simultaneously, peers may need to manage several connections and sets of centroids if they participate in multiple overlays. Solutions such as automatic peer re-election, replication, or fallback references can bolster resilience, but these refinements are outside the immediate scope of this work. 

Currently, no mechanism was chosen to re-balance the tree if too many users have dropped out. This could lead to leaf-nodes being almost entirely depopulated in a section of the tree. Without some kind of re-balancing process happening the tree could tend to become highly imbalanced, thereby negating much of the advantage of the $N \log N$ complexity. Automatic re-balancing by contacting nearby leaf-nodes to check their population size could mitigate this issue, but it was also not part of the scope of this paper.

Lastly, in our experiments, we do not rank the search results. However, we note that ranking can be accomplished locally on the device of the user sending the query (for example, using the cosine-similarity between the query and the retrieved documents from multiple document-providers).

In summary, the decentralized implementation allows each peer to join, leave (with the caveat mentioned above), and expand its neighbor set without depending on a global coordinator or centralized index. By storing centroids at local custodians for each split-node, the network naturally organizes itself into a hierarchical overlay. This structure supports semantic search using the same methods described in our simulations, with only local trust and communication requirements.

\section{Availability}
The AOL4PS data set used~\cite{guo2021aol4ps}, is publicly available from \url{https://doi.org/10.11922/sciencedb.j00104.00093}.
The Python source code that has been used to evaluate this work is publicly available at \url{https://github.com/pneague/Semantica.git}.

\section{Conclusion}
\label{sec:conclusion}

\emph{Semantica} is a fully distributed algorithm for semantic search that utilizes pre-trained large language models (LLMs) to construct a semantic overlay network. By organizing peers into a hierarchical tree based on their document embeddings, Semantica provides a scalable and accurate foundation for decentralized semantic search. Experiments show that Semantica significantly improves peer discovery and document retrieval accuracy over random baselines and graph-diffusion approaches. Our algorithm achieves this superior efficiency through its tree construction and neighbor expansion complexities of \(\mathcal{O}(N \log N)\) and \(\mathcal{O}(r_{\max} \, N \, k)\), respectively, under typical embedding distributions.

Semantica enables effective query routing using chain-hopping, achieving high retrieval rates with minimal communication overhead. The introduction of soft clustering ensures that semantically similar users are well-connected while maintaining network efficiency. Periodic neighbor-expansion rounds further refine user connections, enabling near-complete recall of the most relevant peers.
%
%
Semantica lays a strong foundation for decentralized semantic search, leveraging pre-trained LLM embeddings to create scalable, efficient, and adaptable overlays. With further refinement and extensions, it holds promise as a robust framework for distributed information discovery in large-scale, dynamic networks.

%

\bibliographystyle{IEEEtranS}
\bibliography{bibliography}

\end{document}